\documentclass[twocolumn]{aastex702} 

\usepackage{multirow}

\usepackage{graphics,epsf}
\usepackage[utf8]{inputenc}
\usepackage{amsmath}                
\usepackage{amsfonts}               
\usepackage{amssymb}                
\usepackage{epsfig}                 
\usepackage{graphicx}               
\usepackage{float}
\usepackage{color}
\usepackage{multirow}               

\hypersetup{
    colorlinks=true,
    linkcolor=red,   
    urlcolor=cyan}

\usepackage[colorinlistoftodos]{todonotes}

\newcommand{\cm}{{~\rm cm}}

\newcommand{\s}{{~\rm s}}

\newcommand{\g}{{~\rm g}}
\newcommand{\G}{{~\rm G}}
\newcommand{\K}{{~\rm K}}

\newcommand{\yr}{{~\rm yr}}

\newcommand{\AU}{{~\rm AU}}

\begin{document}

\title{Common Envelope Evolution: Too stochastic to be deterministic}

\author[0000-0003-0375-8987]{Noam Soker}
\affiliation{Department of Physics, Technion - Israel Institute of Technology, Haifa, 3200003, Israel; soker@technion.ac.il}
\email{soker@technion.ac.il}

\begin{abstract}
I argue for significant stochastic behavior in the common envelope evolution (CEE) of cool giants and main-sequence companions to the degree that it is almost impossible to know the exact CEE outcome of an individual binary system from the average properties of the stars and their initial orbit. I start with two observational properties of post-CEE planetary nebulae: many have jet-shaped morphologies, and they show a wide variety of morphologies. The first suggests that the standard CEE model should include jets, and the second shows that the CEE has a wide variety of evolutionary routes; for example, jets can be launched before, during, and/or after the main CEE phase. I show that the properties of the disk with which the main-sequence companion enters the CEE depend sensitively and stochastically on the pulsations and convection of the giant star. Pulsations and giant-envelope convection stochastically determine the evolution of the accretion disk inside the envelope. The stochastic nature of the accretion disk properties implies that jet power is also stochastic. Adopting the view that jets play a major role in CEE outcomes, I conclude that the CEE outcome is stochastic. I conclude that the CEE efficiency parameter $\alpha_{\rm CE}$ has little merit for individual systems, although it still has meaning in population synthesis and other studies of large populations. 
\end{abstract}
   

\section{Introduction} 
\label{sec:intro}

In the common envelope evolution (CEE; e.g., \citealt{Borges2026, DiStefanoetal2026, Grondinetal2026, Gurjaletal2026, Karinoetal2026, LiZetal2026, MuCetal2026, Noughanietal2026, ShiYetal2026, Thaietal2026, Yamaguchietal2026} for some papers from 2026, and \citealt{RoepkeDeMarco2023} for a review), a smaller companion star orbits into the envelope of a giant star. It releases orbital gravitational energy that removes part or all of the giant's envelope. The end state might be a binary system of the giant's core and the companion, or a merger of the companion and the core. One of the major quests of the CEE research is to predict the final orbital separation given the binary properties at the onset of the CEE. 
In this study, I argue that this is a very challenging (almost impossible) task for individual CEE cases with a main-sequence companion because of a significant stochastic component in the CEE, particularly at its initial phase.  

The morphologies of many planetary nebulae (PNe) exhibit opposite (with respect to the center) pairs of lobes, clumps, bubbles, or `ears'; some PNe have one pair, other PNe have two or more pairs along inclined axes, i.e., point-symmetric morphologies. Many studies over the years attributed the shaping of these morphologies to pairs of jets (e.g., \citealt{Morris1987, Soker1990AJ, SahaiTrauger1998, AkashiSoker2018,   EstrellaTrujilloetal2019, Tafoyaetal2019, Balicketal2020, RechyGarciaetal2020, Clairmontetal2022, Danehkar2022, MoragaBaezetal2023, Derlopaetal2024, Mirandaetal2024, Sahaietal2024}; for an alternative idea \citealt{Baanetal2021RAA}). 
Most studies consider a main-sequence companion to launch these jets, as it accretes mass from the asymptotic giant branch (AGB) progenitor of the PN; in some cases the progenitor is a red giant branch (RGB) star (e.g., \citealt{Hillwigetal2017, Sahaietal2017, Jonesetal2020, Jonesetal2022, Jonesetal2023}). 
 
The short orbital periods of binary systems at the center of more than a hundred bipolar and elliptical PNe, post-asymptotic giant branch nebulae,  and pre-PNe (e.g., \citealt{Miszalskietal2019ic, Oroszetal2019, Jones2020Galax, Jones2025}) suggest that they have experienced a CEE phase. In post-AGB and post-RGB systems and PNe with intermediate orbital periods, months to years (e.g., \citealt{VanWinckeletal2014, Moltzeretal2025, VanWinckel2025Galax}), the systems might have experienced the grazing envelope evolution (GEE; e.g., \citealt{Soker2017AGBIB}). 
Many of these have morphologies that are compatible with shaping by pairs of jets. This led me to argue that jets are the most robust observable ingredient of CEE \citep{Soker2025RobustJets}, at least for CEE with main-sequence companions (for other cases with jets, see, e.g., \citealt{Grichener2025}).  
   
In this study, I consider the CEE of binary systems of a giant star with a main-sequence companion; the giant star can be an RGB, an AGB, or a red supergiant star (RSG). During the CEE, these types of cool giants lose mass at a very high rate and form dense dust in the outflow (e.g., \citealt{MuDeMarcoetal2026}); the dust's high opacity might boost the mass loss rate and enhance envelope removal (e.g., \citealt{Soker1998dust, Soker2000dust, Luetal2013, GlanzPerets2018, Iaconietal2019, Iaconietal2020, Reichardtetal2020, BermudezBustamanteetal2024a}; if the dust forms in the already unbound material, then its role in facilitate envelope removal is small, e.g., \citealt{BermudezBustamanteetal2024b}). 
The high opacity of the dust prevents direct observation of the CEE, and most likely of the GEE, with cool giant stars. Studies that compare CEE theory with observations take other approaches besides direct observations. 
  
One common approach to study CEE is to estimate the CEE efficiency parameter, $\alpha_{\rm CE}$ (e.g., \citealt{DeMarcoetal2011, IaconiDeMarco2019, GeHetal2022, GeHetal2024, Grondinetal2026, LiZhenweietal2026, LinJetal2026, ShiYetal2026, ZhangLGeetal2026}). This approach takes the following steps. ($i$) Calculate the energy of an observed post-CEE binary system, $E_{\rm b,f}$. ($ii$) Estimate the binary parameters of its progenitor at the onset of the CEE, and from these, the binary energy at that time, $E_{\rm b,0}$. ($iii$) Calculate the binding energy of the giant's envelope at the onset of the CEE, $E_{\rm bind}$. ($iv$) Calculate the CEE parameter by $\alpha_{\rm CE} \equiv E_{\rm bind}/(E_{\rm b,0}-E_{\rm b,f})$. This definition originated from the traditional CEE (e.g., \citealt{Paczynski1976, Webbink1984}), where the only energy available to remove the giant's envelope in a short time is the orbital energy of the companion-core binary system.
    
The main usage of the $\alpha_{\rm CE}$ prescription is in population synthesis simulations and similar calculations (e.g., \citealt{Rileyetal2022, Yamaguchietal2026}); I accept that this is the best way to conduct population synthesis, i.e., statistical estimates. Other CEE prescriptions exist (e.g., \citealt{Nelemansetal2005, NelemansTout2005, Nelemansetal2025} for a prescription based on angular momentum), but there seems to be no advantage in using these. The SCATTER prescription \citep{DiStefanoetal2023} has problems (e.g., \citealt{CohenSoker2023RAA, Thaietal2026}), and the Two-stage prescription \citep{HiraiMandel2022, Thaietal2026} was criticized by \cite{CohenSoker2023RAA}. 
  
However, the $\alpha_{\rm CE}$ does not incorporate all CEE processes, particularly the launching of jets by the companion. In this study, I argue that the launching of jets by a main-sequence companion introduces a significant stochastic component in the CEE evolution. In Section \ref{sec:Motivation}, I list the assumptions of this study. I will refer mainly to the $\alpha_{\rm CE}$ prescription, and will not deal with other prescriptions nor with numerical hydrodynamical simulations of the CEE. In Section \ref{sec:DiskFormation}, I discuss the formation of a disk inside the envelope; in Section \ref{sec:Entering}, the existence of an accretion disk at the CEE onset; and in Section \ref{sec:DiskEvolution}, the evolution of a disk inside the envelope. All of these processes constitute large stochastic components. I summarize this study in Section \ref{sec:Summary} by claiming that it is hard to predict the outcome of a CEE of an individual binary system of a giant and a main-sequence companion.

\section{Motivation and assumptions} 
\label{sec:Motivation}

I list the observational results and theoretical calculations that motivate me to consider the stochastic nature of the CEE when a main-sequence companion enters the envelope of an RGB, AGB, or RSG star, as well as the main processes that drive this stochastic behavior.   

\textit{(1) The major role of jets.} A key ingredient of the present study is my conclusion from \cite{Soker2025RobustJets} that jets are the most robust observable ingredient of CEE \citep{Soker2025RobustJets} and play significant roles in the CEE, even if active before, during the GEE, or after the CEE. At the beginning of the CEE, the released orbital energy is small, and jets might play the dominant role in removing mass at that phase (e.g., \citealt{Soker2004NewA})

\textit{(2) Wide variety of post-CEE PN morphologies.} 
A reasonable expectation is that if the spiraling-in of the companion towards the core during the CEE and the final orbital separation and timescale were deterministic, so is the ejection of the envelope. However, as \cite{JonesBoffin2017J} nicely noted, there is a wide array of morphologies observed in post-CEE PNe. 
One such property is the launching time of the jets (e.g., \citealt{Guerreroetal2020}). \cite{Tocknelletal2014} estimated that the jets in the PN NGC 6778, which has a post-CEE central binary system (e.g., \citealt{Miszalskietal2011NGC6778, Jonesetal2016}), were launched $\simeq 3000 \yr$ after the CEE, while in ETHOS 1 (binary by \citealt{Miszalskietal2011ETHOS1, Mundayetal2020}), Abell 63 (binary by \citealt{Belletal1994, Afsaretal2008, Corradietal2015}), and the Necklace PN (binary by \citealt{Corradietal2011, Miszalskietal2013}), the jets were launched a few thousand years before the CEE. If jets play a significant role in determining the binary evolution, before, during, or after the CEE, these two different types of cases cannot represent the same evolution. It seems there is a wide variety of envelope-ejection steps in the CEE phase and in the phases before and after it.  

\textit{(3) Main-sequence stars can accrete at a high rate.} To launch energetic jets in a relatively short time, the main-sequence stars should accrete mass at high rates; although the CEE might be long (e.g., \citealt{MichaelyPerets2019, Igoshevetal2020}), some jet-launching episodes might be relatively short (e.g., \citealt{SokerKashi2012, BoumisMeaburn2013}). The larger radii of some main-sequence companions in post-CEE binary systems at the center of PNe indicate that the companion accreted mass during the CEE (e.g., \citealt{Jonesetal2015}).  The accreted material onto the main-sequence star heats up as it slows down near the surface of the main-sequence star, either in a shock if accreted directly, or in the boundary layer of an accretion disk. This high-entropy gas inflates an envelope (e.g., \citealt{SchurmannLanger2024, MukhijaKashi2025a, MukhijaKashi2025b, MukhijaKashi2026a}) that can reduce the accretion rate and prevent accretion disk formation and jet launching. However, one-dimensional (1D) simulations showed that if the jets themselves can remove the high-entropy gas, accretion at a high rate can proceed with very modest expansion of the main-sequence star (e.g., \citealt{BearSoker2025Removal, ScolnicBearSoker2025}).  This is the jetted-mass-removal accretion scenario. In addition, \cite{LjungGilkisTacchella} showed recently that a rapidly rotating main-sequence star can transfer angular momentum to the disk; hence, the rapid rotation of the star does not substantially limit high mass-accretion rates. 

\textit{(4) The significant roles of convection in CEE. } 
Envelope convection plays some roles in the CEE. (a) Most importantly, convection can efficiently transfer outward recombination energy and energy that the companion deposits inside the giant envelope by its orbital decay (e.g., \citealt{Sabachetal2017, WilsonNordhaus2019}), thereby reducing the efficiency of envelope removal (e.g., \citealt{WilsonNordhaus2020, WilsonNordhaus2022, Grondinetal2026}). (b) Another role of the envelope convection is to introduce a stochastic angular momentum component to the mass that the companion star accretes (e.g., \citealt{Dorietal2023, Hilleletal2026}). The other component is due to the density gradient in the envelope and the orbital motion of the companion inside the envelope, and it has a constant direction perpendicular to the orbital plane. These studies found that the two components are of the same order of magnitude, implying that the angular momentum of the accretion disk, if it forms, changes, and the pairs of jets it might launch wobble at large angles and even flip direction. I discuss this further in Section \ref{sec:DiskEvolution}. (c) The large convection cells in cool giants distort the surface of the star (e.g., \citealt{Freytagetal2024, MaJetal2025, Schreieretal2026}). This distorted surface and the pulsation properties influence the mass transfer to the companion, and thereby determine the existence or not of an accretion disk and its properties, when the secondary main-sequence star is outside but very close to the stellar surface (Section \ref{sec:Entering}).  
  
\textit{(5) The formation of a disk around the main-sequence companion in the CEE is marginal.} 
The convective-fluctuation component and the constant-direction component due to the density gradient and orbital velocity are generally not enough to form a Keplerian accretion disk around main-sequence stars that are inside the envelope of giant stars (e.g., \citealt{Dorietal2023, Hilleletal2026}). The idea with a main-sequence companion is that it accretes mass via an accretion disk when outside the envelope (i.e., via Roche-lobe overflow) or when grazing the envelope, and then enters the CEE with an existing accretion disk (e.g., \citealt{Hilleletal2026} and references therein). The specific angular momentum in a Roche lobe overflow is sufficiently large to form an accretion disk, e.g., Table 2 in \cite{LubowShu1975} for analytical calculations with a fit to it by \cite{HessmanHopp1990}, and \cite{JuarezGarciaetal2026} for recent 3D simulations. However, the question is whether the companion has an accretion disk at the onset of the CEE, and how massive it is. As I discuss in Section \ref{sec:Entering}, the properties of the accretion disk and its existence or not at the onset of the CEE introduce a significant stochastic component to the CEE. As stated in Section \ref{sec:intro}, in this study I address the applicability of the $\alpha_{\rm CE}$ CEE parameter and will not consider hydrodynamical simulations of the CEE. 

 \textit{(6) Large-amplitude regular and irregular pulsations.} RSG (e.g., the RSG progenitor of SN~2023ixf, e.g., \citealt{Kilpatricketal2023, Soraisametal2023}) and AGB stars (e.g., \citealt{Trabucchietal2021, Ahmadetal2025}), and to a lesser extent RGB stars, are known to perform oscillations with large amplitudes. In addition to these, there are major expansion phases of these stars, like helium-shell flashes of AGB stars and outbursts of RSG stars. A rapid expansion phase of a giant star might trigger the CEE. The question is then whether an accretion disk exists or not, and if it exists, what is its mass. These are the subject of Section \ref{sec:Entering}. 

The main conclusion from these points and assumptions is that since jets play a large role in mass removal, and since accretion disks launch these jets, whether an accretion disk is formed or not, particularly during the early CEE when the released orbital energy is small, is crucial for the CEE and its outcome.  

\section{An accretion disk inside the envelope} 
\label{sec:DiskFormation}
  
Consider a secondary main-sequence star orbiting inside the envelope of an RGB, AGB,
or an RSG star at an orbital radius $a$. Following earlier studies (e.g., \citealt{Soker2004AM, Dorietal2023, Hilleletal2026}), I assume that the relative velocity of the companion to the envelope is the Keplerian velocity. I also assume that the 3D hydrodynamical simulations of Bondi-Hoyle-Lyttleton (BHL) accretion of a point mass moving linearly in a box by \cite{Livioetal1986} hold for the CEE; the more sophisticated 3D simulations of wind accretion performed by \cite{Kashietal2022} support this assumption. \cite{Livioetal1986} found that the accretion flow becomes non-axisymmetric due to the density gradients, so that the accreted angular momentum is $\eta \simeq 0.2$ times the angular momentum that would have been accreted if the flow had stayed axisymmetric. As in \cite{Hilleletal2026}, I consider the velocity gradient due to the circular orbit, which introduces accreted angular momentum opposite to that from the density gradients, and I scale with $\eta=0.1$, which also incorporates a somewhat lower accretion rate than the BHL one. Namely, it considers that the accretion radius is $\simeq 0.75-1$ times the BHL accretion radius. I also take the density profile in the envelope to be 
\begin{equation}
\rho_{\rm env} (r) \propto r^{-\beta}.
\label{eq:EnvDensty}
\end{equation}
The radius of the accretion disk that this accretion flow forms (see \citealt{Hilleletal2026} for details) is 
\begin{equation}
\begin{split}
 R_{\rm D} = & \eta^2 \beta^2
 \left[ \frac{M_2}{M_1(a)} \right]^{3}  a = 0.52   
\\ \times 
\left( \frac{\eta}{0.1} \right)^2  
&
\left( \frac{\beta}{3} \right)^2 
\left( \frac{M_2}{0.3 M_1(a)} \right)^3 
\left( \frac{a}{1 \AU} \right) R_\odot,
\label{eq:Rdisk2}
\end{split}
\end{equation}
where $M_1(a)$ is the giant's mass inner to the orbit of the secondary star.  

Equation (\ref{eq:Rdisk2}) is a crude one and suffers from several drawbacks. (1) The value of $\eta$ is poorly determined. I scaled (as in previous papers) by results from 3D simulations 40 years ago. Much more sophisticated simulations are needed in a full CEE. (2) The value of $\beta$ is not constant, nor with radius nor with time; i.e., it can vary with time due to pulsations. In the deep envelope of AGB stars, for example, a more reasonable value is $\beta \simeq 2-2.5$, while near the surface $\beta>3$. (3) This formula assumes a spherically symmetric envelope according to equation (\ref{eq:EnvDensty}). This is not the case near the stellar surface, as 3D simulations (e.g., \citealt{HofnerFreytag2019, FreytagHofner2023, Freytagetal2024, MaJetal2025, Schreieretal2026}) show large distortions from a spherical structure; namely, there are non-radial oscillation modes.  

Despite these drawbacks, equation (\ref{eq:Rdisk2}) does show that the formation of the disk is marginal and sensitively depends on several parameters. It is first sensitive to the ratio $M_2/M_1$. More massive stars accrete mass with a higher specific angular momentum, and the disk's radius increases much faster with $M_2$ (for a given $M_1$) than its radius $R_2$. For a given ratio of $M_2/M_1$, the formation of the disk is sensitive to other parameters. These parameters vary stochastically because of the stellar oscillations and envelope convection. I conclude this subsection with the following statement regarding the onset of the CEE, namely, that it occurs just as the secondary main-sequence star enters the photosphere of the cool giant star. \textit{Given two 1D stellar models, a primary red giant (RGB, AGB, or RSG) and a secondary main-sequence star that just enters a CEE, it is impossible to know whether the mass transfer at that given time will lead or not to the formation of an accretion disk around the main-sequence secondary star. } 

\section{Entering the CEE with a disk} 
\label{sec:Entering}

I consider a secondary main-sequence star close to, but outside, the giant envelope, with Roche-lobe-like mass transfer occurring mainly when the cool giant is near its maximum radius in pulsation cycles. 

Analytical calculations (e.g., \citealt{LubowShu1975}) and numerical simulations (e.g.,  \citealt{JuarezGarciaetal2026}) show that large accretion disks form during a Roche lobe overflow from a cool giant to a companion. An accretion disk can easily form around a main-sequence companion. 
 Here I consider the pulsation of the cool giant on a dynamical timescale, $\tau_{\rm p} \simeq 1 \yr$. 
As in \cite{Dorietal2023}, I take the typical disk's lifetime to be $\zeta \simeq 10-100$ times the Keplerian orbital period at its outer radius $R_{\rm D}$  
\begin{equation}
\begin{split}
\tau_{\rm D} &= \zeta \frac{ 2 \pi R_{\rm D}^{3/2}}{(G M_2)^{1/2}}  = 1
\left( \frac{\zeta}{100} \right)   
\\ &    \times
\left( \frac{M_2}{1 M_\odot} \right)^{-1/2}   
\left( \frac{R_{\rm D}}{10 R_\odot} \right)^{3/2}   \yr . 
\label{eq:taudiskD}
\end{split}
\end{equation}
For the above parameter, the disk lifetime is of the order of the pulsation period. If $\tau_{\rm D} \gtrsim \tau_{\rm p}$, the disk survives during the entire pulsation period, while if $\tau_{\rm D} \lesssim \tau_{\rm p}$, it does not, or only a small mass is left in the accretion disk. A disk of little mass will launch, if at all, weak jets. 

Consider that $M_1$ and its average (non-pulsating) envelope structure, $M_2$, and the orbital radius $a$ are given. In the traditional CEE theory, this is sufficient to determine the outcome of the CEE. 
However, the cool giant strongly pulsates, and in one such maximum expansion it might completely engulf the secondary star to onset the CEE. Whether the companion has an accretion disk that launches jets is critical for the CEE outcomes, according to the points of Section \ref{sec:Motivation}. And this, as I claim in this study, is stochastic.
Let me show it here. 

I note that $M_1$ and $a$, which is not much larger than $R_1$, determine the fundamental period $\tau_{\rm p}$. In that case, the ratio $\tau_{\rm D}/\tau_{\rm p}$ is sensitive to the disk radius $R_{\rm D}$ and $\zeta$. The parameter $\zeta$ depends on the disk viscosity: $\zeta$ is larger for lower viscosity. The disk viscosity depends on the magnetic field in the disk. Since the source of the disk material is the giant envelope, the disk magnetic field depends on the initial envelope magnetic fields. Both theoretical expectations (e.g., \citealt{Soker1998}) and observations (e.g., \citealt{Vlemmingsetal2026}, who detected a magnetic flare in an AGB star and confirmed a theoretical prediction by \citealt{SokerKastner2003}) show that AGB magnetic fields can vary with time and with location on the stellar envelope. The magnetic fields are amplified by a dynamo, and therefore their behavior is stochastic. 
The radius $R_{\rm D}$ is also stochastic, because it depends on the accretion flow from the surface of the giant star that faces the secondary star, and its duration. Non-radial pulsation modes (which distort the envelope; e.g., \citealt{HofnerFreytag2019, FreytagHofner2023, Freytagetal2024, MaJetal2025, Schreieretal2026}) introduce stochastic behavior to the accretion process. Namely, the total mass that the cool giant transfers to the secondary star and its specific angular momentum have large stochastic variations.  

To summarize this discussion, the ratio 
\begin{equation}
\frac{\tau_{\rm D}}{\tau_{\rm p}} \approx 
\left[ \frac{ \zeta(\vec{B}_{S})}{100} \right] 
\left[ \frac{ R_{\rm D}(\rm Puls_{S}) }{ 10 R_\odot} \right]^{3/2} F(M_1, M_2, a)  ,  
    \label{eq:Ratio1}
\end{equation}
is largely unpredictable due to several stochastic processes. 
The meaning of these variables is as follows. $\zeta(\vec{B}_{S})$ means that the viscosity depends on the magnetic field, which has stochastic behavior (subscript $S$), $R_{\rm D}(\rm Puls_{S})$ means that the radius of the disk depends on the pulsations of the cool giant, and the subscript $S$ means that the pulsation has stochastic behavior as well. The function $F(M_1, M_2, a)$ is a function of the masses and orbit, and for solar masses and $a= 1 \AU$, $F \approx 1$.  

Going back to the onset of the CEE at the maximum radius in a specific pulsation, the uncertainty in whether the ratio ${\tau_{\rm D}}/{\tau_{\rm p}}>1$ or ${\tau_{\rm D}}/{\tau_{\rm p}}>1$, means we cannot know whether the disk exists or not from the previous maximum radius in the pulsation. 
In addition, and might be more important, the stochastic behavior of the accretion process due to the largely non-spherical stellar surface implies that we cannot know what type of accretion disk is formed at the specific pulsation that triggers the CEE. Namely, what its mass is and what direction its angular momentum axis is

\section{Disk evolution during early CEE} 
\label{sec:DiskEvolution}

\subsection{The accretion rate} 
\label{subsec:AccretionRate}

Consider a main-sequence secondary star inside the envelope accreting at the BHL rate under the assumptions stated above: it moves supersonically through the giant envelope at its Keplerian velocity, and I neglect the sound speed in the calculation of the accretion rate.
Scaling with typical cool giant values in the outer envelope (not too close to the core), yields an accretion rate of  
\begin{equation}
\begin{split}
\dot M_{\rm 2,BHL} & \simeq 1.2 
\left( \frac{\rho_{\rm env}}{10^{-7}\g \cm^{-3}} \right)^2 
\left( \frac{M_2}{0.3 M_1(a)} \right)^2 
\\ & \times 
\left( \frac{a}{1 \AU} \right)^{3/2}
\left( \frac{M_1}{1 M_\odot} \right)^{1/2} 
M_\odot \yr^{-1} .
\label{eq:RBHL}
\end{split}
\end{equation}
  
If a companion launches jets in the giant envelope, the jets remove and heat the envelope, and by doing so reduce the density in the companion vicinity; this, in turn, reduces the accretion rate. The process is a negative CEE jet feedback cycle (see, e.g., \citealt{Soker2016Rev} for a review, \citealt{GrichenerCohenSoker2021} for 1D simulations, and \citealt{Hilleletal2022} for 3D simulations). \cite{WeinerSoker2025} find for a low-mass main-sequence star in a CEE with an AGB star that the jets launched by a main-sequence companion reduce the density in the companion vicinity by a factor of $\chi \approx 0.5 (M_2/0.1 M_\odot)^{-1}$. For $M_2=0.5 M_\odot$ this crude fitting gives $\chi \approx 0.1$. 

I approach the accretion rate from a different direction. The potential well of a main-sequence star is $\approx 100$ times as deep as the AGB potential well in the envelope. To release accretion energy about equal to the binding energy of the outer envelope, the companion should accrete $\approx 0.01$ of the envelope mass. The jets carry a fraction of this energy. For an AGB envelope mass of $\approx 1 M_\odot$ and a spiraling-in time about equal the Keplerian time, $\tau_{\rm p} \simeq 1 \yr$, the accretion rate should be then $\dot M_{\rm 2, acc} \simeq 0.01-0.1 M_\odot \yr^{-1}$.  I will scale with this accretion rate in what follows.    
 
\subsection{The disk density} 
\label{subsec:DiskDensity}
 
The density in the accretion disk around the companion at radius $r_d$ from the center of the secondary star for a relaxed thin accretion disk is (e.g., equation 5.49 in \citealt{FrankKingRaine2002})  
\begin{equation}
\begin{split}
\rho_d & \simeq  8.5 \times 10^{-7} 
\left( \frac{ \alpha_{\rm d}}{0.1} \right) ^{-7/10}
\left( \frac {\dot M_{\rm 2, acc}}{10^{-2} M_\odot \yr^{-1}}    \right)^{11/20} 
\\ & \times
\left( \frac {M_2}{1 M_\odot} \right)^{5/8} 
\left( \frac {r_d}{5 R_\odot} \right)^{-15/8} 
\left( \frac {f}{0.5} \right) ^{11/5} \g \cm^{-3} ,
\label{eq:RhoDisk}
\end{split}
\end{equation}
where $\alpha_{\rm d}$ is the Shakura-Sunyaev viscosity parameter, and $f$ is given by 
\begin{equation}
f = \left(1 - \sqrt{\frac{R_2}{r_d}}\right)^{\frac{1}{4}}  .
\label{eq:f}
\end{equation}
I scaled with a radius in the disk much larger than the stellar radius for a solar-mass main-sequence star, and with an accretion rate much lower than the BHL one. Despite these conservative values, the density in the disk is an order of magnitude above that of the cool giant envelope. The conclusion is that when accretion occurs through an accretion disk in a CEE, the disk density is much larger than that of the cool envelope; hence, it can withstand the ram pressure as it moves with the companion inside the giant envelope.  

The question then is the specific angular momentum of the accreted gas. 

\subsection{Disk survivability} 
\label{subsec:DiskSurvivability} 
\subsubsection{Disk angular momentum} 
\label{subsubsec:AMconsiderations} 
I assume that the secondary star enters the giant envelope with an already existing accretion disk. I consider a thin accretion disk, with height $H(r_d) \ll r_d$.  The disk density is $\rho_d(r_d)$, its surface density is $\Sigma (r_d) = 2 H \rho_d(r_d)$, and it extends from the secondary radius $R_2$ to an outer radius $R_{\rm D}$. The total angular momentum in the accretion disk is 
\begin{equation}
\begin{split}
J_{\rm D,0} & =\int^{R_{\rm D}}_{R_{\rm 2}} 2 \pi r_d dr_d \Sigma \sqrt{G M_{\rm 2} r_d} = 
M_{\rm D,0} j_2 
\\ &\times 
 \left[ \frac{5}{7}\frac{(R_{\rm D}/R_2)^{7/4} - 1}{(R_{\rm D}/R_2)^{5/4} - 1}  \right] \equiv M_{\rm D,0} j_2  Q,
    \label{eq:JDisk}
\end{split}
\end{equation}
where $M_{\rm D,0}$ is the accretion disk mass at that time, and $j_2=\sqrt{G M_2 R_2}$ is the specific angular momentum in the inner disk region that launches the jets. The third equality defines $Q$ to be the factor inside the square parenthesis: $Q=2.35$ for $R_{\rm D}=10 R_2$, $Q=1.73$ for $R_{\rm D}=5 R_2$, $Q=1.32$ for $R_{\rm D}=2.5 R_2$, and $Q=1$ for $R_{\rm D} \rightarrow R_2$.

After entering the CEE with an accretion disk, the secondary star accretes mass from teh envelope. The accretion disk will survive as long as the specific angular momentum of the disk is $J_{\rm D}/M_{\rm D}>j_2$.  The accreted gas inside the envelope has an angular momentum that originates from two sources (Section \ref{sec:DiskFormation}): (a) the constant-axis angular momentum source due to the density gradient in the giant envelope and the orbital motion; its specific angular momentum is $j_{\rm O}$; (b) the stochastic one due to the giant envelope random convection motion; its specific angular momentum in $j_{\rm R}(t)$ and it depends stochastically on time. If the first source of specific angular momentum is larger than $j_2$, namely, $R_{\rm D}>R_2$ (equation \ref{eq:Rdisk2}), the disk will continue to survive for a long time. If it is smaller, the disk might slowly be depleted. Below I will consider the accretion of gas with $j_{\rm O}<j_2$

In principle, the accretion disk can live forever, as viscosity transfers angular momentum outward inside the disk, and material with zero angular momentum is accreted through the disk and onto the secondary star. This is, of course, not accurate, as the central object, here a main-sequence star, rotates and its radius is non-negligible, so the final angular momentum of the accreted gas is not zero and is non-negligible. The jets also carry away angular momentum. I assume that the jets and secondary rotation remove angular momentum at the inner disk radius, at the secondary surface. This gas has a specific angular momentum of $j_2=\sqrt{G M_2 R_2}$. 

\subsubsection{The constant-axis angular momentum} 
\label{subsubsec:ConstantAM} 

Consider that the convective angular momentum averages to zero over the lifetime of the accretion disk (the other limit is in Section \ref{subsubsec:RandomAM}). This is the case if the accretion disk is relatively large, a disk radius of $R_{\rm D} \gtrsim 5 R_\odot$. 
Over a long time, the average specific angular momentum of the accreted gas is $j_{\rm O}$. As stated above, if $j_{\rm O} > j_2$ the disk will survive for a long time. However, as equation (\ref{eq:Rdisk2}) shows, in many cases $j_{\rm O} < j_2$, with large uncertainties. Even in that case, the accreted angular momentum is not negligible, and the expectation is that
\begin{equation}
k_j \equiv \frac{j_{\rm O}}{j_2} \simeq 0.1-1.  
\label{eq:jRatio}
\end{equation}

The secondary star accretes an extra mass $\Delta M_{\rm D}$ while inside the envelope. 
The total angular momentum that is lost at the inner radius until the depletion of the disk, to the accreted mass and jets, is $(M_{\rm D,0} + \Delta M_{\rm D}) j_2$. This equals the initial angular momentum of the disk $J_{\rm D,0}$, and the accreted one $\Delta M_{\rm D} k_j j_2 $. The equality reads, using  equation (\ref{eq:JDisk}),  
\begin{equation}
  (M_{\rm D,0} + \Delta M_{\rm D}) j_2 = \Delta M_{\rm D} k_j j_2  + M_{\rm D,0} j_2  Q. 
\label{eq:AMequality}
\end{equation}
Equation (\ref{eq:AMequality}) gives the extra mass until the disk ceases to exist under the above assumptions 
\begin{equation}
\Delta M_{\rm D} = M_{\rm D,0} \frac{Q-1}{1-k_j}.
    \label{eq:DeltaM}
\end{equation}
As the specific angular momentum of the accreted mass increases, i.e., $k_j \rightarrow 1$, the extra mass in the disk increases. A crude typical value is  $\Delta M_{\rm D} \approx M_{\rm D,0}$.

The above estimate is non-negligible. For example, take $R_{\rm D}= 10 R_\odot$ in equation (\ref{eq:taudiskD}) for the disk viscosity time, i.e, $ \tau_{\rm D} \simeq 1 \yr$. For an accretion rate at the entrance to the CEE of $10^{-2} M_\odot \yr^{-1}$, the mass in the disk is $\approx 0.01 M_\odot$. For $Q\simeq 2$ and $k_j \simeq 0.5$, equation (\ref{eq:DeltaM}) gives $\Delta M_{\rm D} \approx 0.02 M_\odot$. In total, the companion accretes $\approx 0.03 M_\odot$. Since the gravitational well of the main-sequence companion is $\approx 100$ times as deep as that of the AGB outer envelope, this energy is equivalent to the binding energy of $\approx 3 M_\odot$. The jets take only a fraction of this energy. The rest is carried by the convection of the giant envelope outward. The conclusion is that the jets that a main-sequence companion might launch in the outer envelope might play a significant role in helping envelope removal. 

The point of this study is that the stochastic nature of different parameters makes an individual prediction impossible. Statistical averaging is possible, but not a deterministic prediction for individual cases.  

\subsubsection{The stochastic convective angular momentum} 
\label{subsubsec:RandomAM} 

The assumption in deriving equation (\ref{eq:DeltaM}) was that the fluctuation of the angular momentum of the accreted mass due to giant envelope convection averaged to zero over the viscous time scale of the disk. However, this is marginal. 
Simple 1D estimates (e.g., \citealt{Dorietal2023}) and 3D simulations (e.g., \citealt{Hilleletal2026}) show two important properties of the angular momentum fluctuations due to giant envelope convection: (a) The specific angular momentum amplitude can be larger by a factor of a few than the constant-axis angular momentum components, $j_{\rm R} \approx {\rm few} \times j_{\rm O}$, and (b) the variation time is $\tau_{\rm R} \approx 0.05-0.1 \yr$. If the accretion disk is small, say $R_{\rm D} \simeq 3 R_{\odot}$ for a solar mass companion, then the disk lifetime is of the order of two months, $\tau_{\rm D} \simeq 0.1 -0.2\yr$. This is not much longer than the fluctuation time scale of the stochastic angular momentum component. In turn, this implies that a large angular momentum fluctuation with an angular momentum axis in the opposite sense to that of the disk can destroy the disk by reducing its angular momentum to below $j_2$. 
Of course, a large angular momentum fluctuation in the same sense as that of teh disk will increase the disk mass, radius, and lifetime.  

The bottom line of this short subsection is that in many cases at the onset of the CEE the secondary star might have a disk with $\tau_{\rm D} \approx \tau_{\rm R}$, and large fluctuations in the angular momentum of the accreted mass can make a long-lived disk, destroy it in a short time, or anything in between. The fate of the accretion disk is stochastic.  

\section{Summary}  
\label{sec:Summary}

I presented arguments for significant stochastic behavior in the CEE of a cool giant (RGB, AGB, or RSG) and its main-sequence companion. The conclusion is that it is almost impossible to know the exact CEE outcome of an individual system from the average properties of the stars and their initial orbit. There is still a meaning to the average outcomes of many systems, and, therefore, to the usage of the CEE efficiency parameter $\alpha_{\rm CE}$ in population synthesis and other studies of large populations. However, using $\alpha_{\rm CE}$ for individual binary systems is much less valuable and cannot predict the outcome.

In reaching this conclusion, I started with two observable properties of post-CEE PNe (points 1 and 2 in Section \ref{sec:Motivation}, where other points and assumptions are listed): (1) Many of the post-CEE PNe are shaped by jets; hence jets play a major role in the CEE \citep{Soker2025RobustJets}. The companion launches these jets. (2) The wide variety of morphologies of post-CEE PNe \citep{JonesBoffin2017J} suggests that the CEE has a wide variety of processes, and there is no single evolutionary route of the CEE. For example, jets can be launched before or after the main CEE phase (e.g., \citealt{Tocknelletal2014, Guerreroetal2020}).  

In Section \ref{sec:DiskFormation}, I pointed out that the formation of an accretion disk due to the density gradient in the envelope and the orbital motion is marginal. Namely, the specific angular momentum of this angular momentum component, $j_{\rm O}$, marginally forms an accretion disk around a main-sequence star: the accretion disk radius it yields is similar to the main-sequence star radius (equation \ref{eq:Rdisk2}). This crude equality implies that the disk may or may not form, depending on stochastic behavior in the giant envelope, particularly pulsations and convective motion. If it forms, it might be short-lived. The scenario with main-sequence companions is that they enter the CEE with an already existing accretion disk. 

When the main-sequence companion is outside the giant envelope but close, Roche-lobe overflow accretion forms an accretion disk (e.g., \citealt{JuarezGarciaetal2026}). However, stochastic processes also occur here, as I discussed in Section \ref{sec:Entering}. Because of the cool giant's large-amplitude pulsations, high-rate mass transfer, such as Roche-lobe-like mass transfer, begins at the maximum pulsation radius. I pointed out that the viscosity lifetime of an accretion disk (equation \ref{eq:taudiskD}) is of the order of the pulsation period (equation \ref{eq:Ratio1}). The disk properties depend on the pulsations, which are irregular and include non-radial modes. Therefore, in some cases the accretion disk will be more massive than in others. It is impossible to know the exact properties of such disks, the radius and mass that remain, if any, for the next mass-transfer episode. Therefore, it is impossible to know the exact disk mass and radius when the secondary star eventually enters a full CEE.   

In Section \ref{sec:DiskEvolution},  I discussed the evolution of the disk after the secondary star enters the CEE. The disk density is high enough (equation \ref{eq:RhoDisk}) to withstand the ram pressure of the giant envelope as the secondary orbits inside it. In Section \ref{subsubsec:ConstantAM} I consider cases where the fluctuations of angular momentum due to the envelope convection average to zero on a time shorter than the disk lifetime. The angular momentum of the accreted gas arises from the density gradient in the envelope and orbital motion, and it has a constant-axis angular momentum along the binary one. I showed that accretion can prolong the disk lifetime and the total mass accreted through the disk (equation \ref{eq:DeltaM}), and thus the jet power. This power can be small or significant, again depending on specific properties of the accretion disk at the envelope entrance and on pulsation and convection in the giant envelope.  
 
I note here the negative CEE jet feedback cycle (Section \ref{subsec:AccretionRate}), in which the jets regulate their power; particularly, very powerful jets can reduce the accretion rate, hence the jets' power. This negative cycle moderates the CEE stochastic nature somewhat. 

Throughout the paper, I emphasize the large stochastic role of convection. (a) Convection distorts the surface of the pulsating giant star; hence, it influences mass transfer to the companion when still outside the envelope. (b) The convection introduces very large angular momentum fluctuations of the accreted gas, which in many cases are several times that of the fixed-axis angular momentum and cause wobbling of the angular momentum axis, hence the jet axis \citep{Dorietal2023, Hilleletal2026}. In Section \ref{subsubsec:RandomAM}, I pointed out that such large fluctuations can substantially prolong the lifetime of the accretion disk in the envelope, or, in the other extreme, can destroy it in a short time. (c) The lifetime of the accretion disk, hence its mass, depends on the viscosity, which depends on the magnetic field. Since the disk material originates in the giant envelope, the initial magnetic field depends on the giant envelope's magnetic field. The magnetic field in the giant envelope is expected to vary stochastically with time and location, as in dynamos of convective envelopes.

In arguing for stochastic CEE behavior with main-sequence companions, I adopted the view that the standard CEE model should include jets, unlike the traditional one that does not. However, this view is not in consensus, and it remains one of several major open questions in the CEE process.

\section*{Acknowledgements}
A grant from the Pazy Foundation 2026 supported this research.
I thank the Charles Wolfson Academic Chair at the Technion for the support.





 \bibliography{BibReference}{}
\bibliographystyle{aasjournalv7.1}

\end{document}